\documentclass[journal,10pt,twocolumn]{IEEEtran}

\ifCLASSINFOpdf
 \else
\fi
\usepackage{amsmath}
\usepackage{hyperref}
\usepackage{graphicx}      
\usepackage{psfrag}
\usepackage[numbers]{natbib}        
\usepackage{amssymb}
\usepackage{amsmath}
\usepackage{dsfont}
\usepackage{euscript}
\usepackage{color}
\usepackage[normalem]{ulem}
\usepackage{overpic}
\usepackage{autobreak}
\allowdisplaybreaks
\newtheorem{definition}{Definition}
\newtheorem{cor}{Corollary}
\newtheorem{lem}{Lemma}
\newtheorem{rem}{Remark}

\newtheorem{theorem}{Theorem}
\newtheorem{assumption}{Assumption}
\newcommand{\E}{\mathrm{E}}

\newcommand{\R}{\mathrm{Re}}

\def\R{\mathbb{R}}

\def\e{\epsilon}
\def\Z{\mathbb{Z}}

\def\xb{\mathbf{x}}

\def\zbx{\mathbf{X}}
\def\ab{\mathbf{a}}

\def\yb{\mathbf{y}}
\def\bmu{\boldsymbol{\mu}}
\def\be{\boldsymbol{\eta}}

\def\Mb{\boldsymbol{M}}
\def\Qb{\boldsymbol{Q}}

\def\Gb{\boldsymbol{G}}

\def\Rb{\boldsymbol{R}}
\def\Ab{\boldsymbol{A}}

\def\gb{\mathbf{g}}
\def\R{\mathbb{R}}

\def\Z{\mathbb{Z}}

\def\xb{\mathbf{x}}

\def\zbx{\mathbf{X}}
\def\ab{\boldsymbol{a}}

\def\by{\boldsymbol{y}}

\def\bx{\boldsymbol{x}}
\def\Proj{\mbox{Proj}}

\def\r{~}

\begin{document}

\title{Learning Nash Equilibria in Monotone Games}

\author{Tatiana~Tatarenko and Maryam Kamgarpour, IEEE Member
\thanks{T. Tatarenko (\url{tatiana.tatarenko@rmr.tu-darmstadt.de}) is with the Control Methods and Robotics Lab Technical University Darmstadt, Darmstadt, Germany 64283, M. Kamgarpour (\url{maryamk@ethz.ch}) is with the Automatic Control Laboratory, ETH Z\"urich, Switzerland.}
\thanks{ M. Kamgarpour gratefully acknowledges ERC Starting Grant CONENE.}
}

\maketitle

\begin{abstract}
We consider multi-agent decision making where each agent's cost function depends on all agents' strategies. We propose a distributed algorithm to learn a Nash equilibrium, whereby each agent uses only obtained values of her cost function at each joint played action, lacking any information of the functional form of her cost or other agents' costs or strategy sets. In contrast to past work where convergent algorithms required strong monotonicity,  we prove algorithm convergence under mere monotonicity assumption. This significantly widens algorithm's applicability, such as to games with linear coupling constraints. 

\end{abstract}


\section{Introduction}

Game theory is a powerful framework for analyzing and optimizing multi-agent decision making problems. In several such problems, each agent (referred to  also as a player) does not have full information on her objective function, due to the unknown interactions and other players' strategies affecting her objective. Consider for example, a transportation network in which an agent's objective is minimizing travel time or an electricity network in which an agent's objective is minimizing own's electricity prices. In these instances, the travel times and prices, respectively, depend non-trivially on the strategies of other agents. Motivated by this limited information setup, we consider computing Nash equilibria given only the so-called \emph{payoff-based information}. That is, each player can only observe the values of its objective function at a joint played action, does not know the functional form of her or others' objectives, nor the strategy sets and actions of other players, and cannot communicate 
with other players. In this setting, we address the question of how  agents should update their actions to converge to a Nash equilibrium strategy. 
 
A large body of literature on learning Nash equilibria with  payoff-based information has focused on finite action setting or potential games, see for example, \cite{pradelski2012learning, Shamma2015,MardRev} and references therein. For games with continuous (uncountable) action spaces, a payoff-based approach was developed based on the extremum seeking idea in optimization \cite{frihauf2012nash,stankovic2012distributed}, and assuming strongly convex objectives almost sure convergence to the Nash equilibrium was proven. A  payoff-based approach, inspired by the \emph{logit} dynamics in finite action games \cite{blume1993statistical} was extended to continuous action setting for the case of potential games  \cite{Tat_cdc16}. The work in  \cite{ZhuFrazzoli} considered learning Nash equilibria in continuous action games on networks. Crucially, the work  additionally assumed that each player exchanges information with her neighbors, to facilitate estimation of the gradient of her  objective function online. 

Recently, we proposed a payoff-based approach to learn Nash equilibria in a class of convex games \cite{TAC2018tat_kam}. Our approach hinged upon  connecting Nash equilibria of a game to the solution set of a related variational inequality problem. Our algorithm convergence was established for the cases in which the game mapping is strongly monotone or the game admits a potential function. Apart from possibly limited scope of a potential game, strong monotonicity can be too much to ask for. In particular, if the objective function of an agent is linear in her action or in the presence of  coupling constraints of the action sets the game mapping will not be strongly monotone. 

Our goal here is to extend the existing payoff-based learning approaches to a broader class of games characterized by monotone game mappings. While algorithms for solving monotone variational inequalities exist (see, for example, Chapter 12 in \cite{FaccPang1}), these algorithms either consist of two 
timescales (Tikhonov regularization approach) or have an extra gradient step (extra-gradient methods). As such, they require more coordination between players than that possible in a payoff-based only information structure. 

Our contributions are as follows. First, we propose a distributed payoff-based algorithm to learn Nash equilibria in a monotone game, extending our past work \cite{TAC2018tat_kam} applicable to strongly monotone games, inspired by the  single timescale algorithm for solving stochastic variational inequalities  \cite{koshal2010single}. Second, despite lack of gradients in a payoff-based information, contrary to the setup in \cite{koshal2010single}, we show that our proposed procedure can be interpreted as a stochastic gradient descent with an additional biasL and regularization terms. Third, we prove convergence of the proposed algorithm to Nash equilibria by suitably bounding the bias and noise variance terms using established results on boundedness and convergence of discrete-time Markov processes.  

\textbf{Notations.} The set $\{1,\ldots,N\}$ is denoted by $[N]$. Boldface is used to distinguish between vectors in a multi-dimensional space and scalars.
Given $N$ vectors $\bx^i\in\R^d$, $i\in[N]$, $(\bx^i)_{i=1}^{N}:=({\bx^1}^{\top}, \ldots, {\bx^N}^{\top})^{\top} \in \R^{Nd}$; $\bx^{-i}:=({\bx^1}, \ldots, {\bx^{i-1}},{\bx^{i+1}}, \ldots, {\bx^N}) \in \R^{(N-1)d}$.
$\R^d_{+}$ and $\Z_{+}$ denote respectively, vectors from $\R^d$ with non-negative coordinates and non-negative whole numbers.  The standard inner product on $\R^d$ is denoted by $(\cdot,\cdot)$: $\R^d \times \R^d \to \R$, with associated norm $\|\bx\|:=\sqrt{(\bx, \bx)}$.  Given some matrix $A\in\R^{d\times d}$, $A\succeq(\succ)0$, if and only if $\bx^{\top}A\bx\ge(>)0$ for all $\bx\ne 0$. We use the big-$O$ notation, that is, the function $f(x): \R\to\R$ is $O(\gb(x))$ as $x\to a$, $f(x)$ = $O(g(x))$ as $x\to a$, if $\lim_{x\to a}\frac{|f(x)|}{|g(x)|}\le K$ for some positive constant $K$.
 We say that a function $f(\bx)$ grows not faster than a function $g(\bx)$ as $\bx\to\infty$, if there exists a positive constant $Q$ such that $f(\bx)\le g(\bx)$ $\forall \bx$ with $\|\bx\|\ge Q$.

\begin{definition}\label{def:mm}
A mapping $\Mb:\R^d\to\R^d$ is \emph{monotone} over $X\subseteq\R^d$, if $(\Mb(\bx)-\Mb(\by),\bx-\by)\ge 0$ for every $\bx,\by\in X$.
 \end{definition}

\section{Problem Formulation}\label{sec:problem}
Consider a game $\Gamma (N, \{A_i\}, \{J_i\})$ with $N$ players, the sets of players' actions $A_i\subseteq \R^d$, $i\in[N]$, and the cost (objective) functions $J_i:\Ab\to\R$, where $\Ab = A_1\times\ldots\times A_N$ denotes the set of joint actions. We restrict the class of games as follows.

\begin{assumption}\label{assum:convex}
 The game under consideration is \emph{convex}. Namely, for all $i\in[N]$ the set $A_i$ is convex and closed, the cost function $J_i(\ab^i, \ab^{-i})$ is defined on $\R^{Nd}$, continuously differentiable in $\ab$ and convex in $\ab^i$ for  fixed $\ab^{-i}$.
\end{assumption}

\begin{assumption}\label{assum:CG_grad}
The mapping $\Mb:\R^{Nd}\to\R^{Nd}$, referred to as the \emph{game mapping}, defined by
 \begin{align}\label{eq:gamemapping}
 \nonumber
 \Mb(\ab) &= (\nabla_{\ab^i} J_i(\ab^i, \ab^{-i}))_{i=1}^N=(\Mb_1(\ab), \ldots, \Mb_N(\ab))^{\top},\\ \nonumber
 &\mbox{where }\Mb_i(\ab) = (M_{i,1}(\ab), \ldots, M_{i,d}(\ab))^{\top}, \mbox{ and}\\
 M_{i,k}(\ab)&= \frac{\partial J_i(\ab)}{\partial a^i_k},\quad \ab\in\Ab, \quad i\in[N], \quad k\in[d],
 \end{align}
 is \emph{monotone on $\Ab$} (see Definition \ref{def:mm}).
\end{assumption}

We consider a \emph{Nash equilibrium} in game $\Gamma (N, \{A_i\}, \{J_i\})$  as a stable solution outcome because it represents a joint action from which no player has any incentive to unilaterally deviate.

\begin{definition}\label{def:NE}
 A point $\ab^*\in\Ab$ is called a \emph{Nash equilibrium} if for any $i\in[N]$ and $\ab^i\in A_i$
 $$J_i(\ab^{i*},\ab^{-i*})\le J_i(\ab^{i},\ab^{-i*}).$$
 \end{definition}
Our goal is to learn such a stable action in a game through designing a payoff-based algorithm. We first connect existence of Nash equilibria for $\Gamma (N, \{A_i\}, \{J_i\})$ with solution set of a corresponding variational inequality problem.
\begin{definition}
Consider a mapping $\boldsymbol T(\cdot)$: $\R^d \to \R^d$ and a set $Y \subseteq \R^d$. A
\emph{solution $SOL(Y,\boldsymbol T)$ to the variational inequality problem} $VI(Y,\boldsymbol T)$ is a set of vectors $\yb^* \in Y$ such that $(\boldsymbol T(\yb^*), \yb-\yb^*) \ge 0$, $\forall \yb \in Y$.
\end{definition}

\begin{theorem}\label{th:VINE}(Proposition\r1.4.2 in \citep{FaccPang1})
 Given a game $\Gamma (N, \{A_i\}, \{J_i\})$ with game mapping $\Mb$, suppose that the action sets $\{A_i\}$ are closed and convex, the cost functions $\{J_i\}$ are continuously differentiable in $\ab$ and convex in $\ab^i$ for every fixed $\ab^{-i}$ on the interior of $\Ab$. Then, some vector $\ab^*\in \Ab$ is a Nash equilibrium in $\Gamma$, if and only if $\ab^*\in SOL(\Ab,\boldsymbol M)$.
\end{theorem}

It follows that under Assumptions~\ref{assum:convex} and \ref{assum:CG_grad} for a game with mapping $\boldsymbol M$, any solution of $VI(\Ab,\boldsymbol M)$ is also a Nash equilibrium in such games and vice versa. 
While $\Gamma (N, \{A_i\}, \{J_i\})$ under Assumptions~\ref{assum:convex} and \ref{assum:CG_grad} might admit a Nash equilibrium, these two assumptions alone do not guarantee existence of a Nash equilibrium. To guarantee existence, one needs to consider a more restrictive assumption, for example, strong monotonicity of the game mapping  or compactness of the action sets \citep{FaccPang1}.  Here, we do not restrict our attention to such cases. However, to have a meaning discussion, we do assume existence of at least one Nash equilibrium in the game.
\begin{assumption}\label{assum:existence}
 The set $SOL(\Ab,\Mb)$ is not empty.
\end{assumption}

\begin{cor}\label{cor:existence}
  Let $\Gamma (N, \{A_i\}, \{J_i\})$ be a game with game mapping $\Mb$ for which Assumptions~\ref{assum:convex}, \ref{assum:CG_grad}, and \ref{assum:existence} hold. Then, there exists at least one Nash equilibrium in $\Gamma$. Moreover, any Nash equilibrium in $\Gamma$ belongs to the set $SOL(\Ab,\Mb)$. 
 \end{cor}

The following additional assumptions are needed for convergence of the proposed payoff-based algorithm to a Nash equilibrium (see proofs of Lemma~\ref{lem:boundedvec} and Theorem~\ref{th:main}). 
\begin{assumption}\label{assum:Lipschitz}
 Each element $\Mb_i$ of the game mapping $\Mb:\R^{Nd}\to\R^{Nd}$, defined in Assumption \eqref{assum:CG_grad} is Lipschitz continuous on $\R^{d}$ with a Lipschitz constant $L_i$.
\end{assumption} 


\begin{assumption}\label{assum:inftybeh}
  Each cost function $J_i(\ab)$, $i\in[N]$, grows not faster than a linear function of $\ab$ as $\|\ab\|\to\infty$.
\end{assumption}


\section{Payoff-Based Algorithm}\label{sec:analysis}


Given a payoff-based information, each agent has access to its current action, referred to as its state and denoted by $\xb^i(t)=(x^i_1,\ldots,x^i_d)^{\top}\in\R^d$, and the cost value $\hat J_i(t)$ at the joint states $\xb(t)=(\xb^1(t),\ldots,\xb^N(t))$, $\hat J_i(t) =J_i(\xb(t)) =J_i(\xb^1(t),\ldots,\xb^N(t))$ at iteration $t$. Using this information in the proposed algorithm each agent $i$ ``mixes'' its next state $\xb^i(t+1)$. Namely, it chooses $\xb^i(t+1)$ randomly according to the multidimensional normal distribution $\EuScript N(\bmu^i(t+1)=(\mu^i_1(t+1),\ldots,\mu^i_{d}(t+1))^{\top},\sigma(t+1))$ with the density:
\begin{align*}
 p_i&(x^i_1,\ldots,x^i_{d};\bmu^i(t+1),\sigma(t+1))\cr
 & = \frac{1}{(\sqrt{2\pi}\sigma(t+1))^{d}}\exp\left\{-\sum_{k=1}^{d}\frac{(x^i_k-\mu^i_k(t+1))^2}{2\sigma^2(t+1)}\right\}.
\end{align*}
The initial value of the means $\bmu^i(0)$, $i  \in \{N \}$, can be set  to any finite value. The successive means are updated as follows:
 \begin{align}\label{eq:regpl}
 &\bmu^i (t+1)=\Proj_{A_i}\big[\bmu^i(t) \cr
 &-\gamma(t)\sigma^2(t)\left({\hat J_i(t)}\frac{{\xb^i(t)} -\bmu^i(t)}{\sigma^2(t)} + \epsilon(t)\bmu^i(t)\right)\big].
 \end{align}
 In the above, $\Proj_{C}[\cdot]$ denotes the projection operator on set $C$, $\gamma(t)$ is a step-size parameter and $\epsilon(t)>0$ is a  regularization parameter. We highlight the difference between the proposed approach and that of \cite{TAC2018tat_kam} due to the additional term $\epsilon(t)$ in \eqref{eq:regpl}. In the absence of this term the algorithm would not be convergent under a mere monotonicity assumption on the game mapping (see counterexample provided in \cite{grammatico2018comments}). 
 
Let us provide insight into the algorithm  by deriving an analogy to a regularized stochastic gradient algorithm. 
Given $\sigma > 0$, for any $i\in[N]$ define $ \tilde{J}_i : \R^{Nd} \rightarrow \R$ as
 \begin{align}
 \label{eq:mixedJ}
  \tilde{J}_i &(\bmu^1,\ldots,\bmu^N, \sigma)= \int_{\mathbb R^{Nd}}J_i(\bx)p(\bmu, \bx, \sigma)d\bx,  
 \end{align}
where $p(\bmu, \bx, \sigma)=\prod_{i=1}^Np_i(x^i_1,\ldots,x^i_{d};\bmu^i,\sigma)$. 
Above, $\tilde{J}_i$, $i\in[N]$, can be interpreted as the $i$th player's cost function in mixed strategies. We can now show that the second term inside the projection in \eqref{eq:regpl} is a sample of the gradient of this cost function $\tilde{J}_i$ with respect to the mixed strategies. Let $\bmu(t) =(\bmu^1(t),\ldots,\bmu^N(t))$.
\begin{lem} 
\label{lem:sample_grad}
Under Assumptions\r\ref{assum:convex} and\r\ref{assum:inftybeh}, $\forall i\in[N], k\in[d]$
\begin{align}\label{eq:mathexp}\nonumber
& \frac{\partial {\tilde J_i(\bmu(t), \sigma(t))}}{\partial \mu^i_k}=\E_{\xb(t)}\{\hat J_i(t)\frac{x^i_k(t) -\mu^i_k(t)}{\sigma^2(t)}\} \\ \nonumber
 =&\E\{J_i(\xb^1(t),\ldots,\xb^N(t))\frac{x^i_k(t) -\mu^i_k(t)}{\sigma^2(t)}|\\
 &\qquad\qquad x^i_k(t)\sim\EuScript N(\mu_k^i(t),\sigma(t)), i\in[N], k\in[d]\}.
 \end{align}
\end{lem}

\begin{IEEEproof}
We  verify that  the differentiation under the integral sign in \eqref{eq:mixedJ} is justified. It can then readily be verified that  \eqref{eq:mathexp} holds, by taking the differentiation inside the integral. A sufficient condition for differentiation under the integral is that the integral of the formally differentiated function with respect to $\mu^i_k$  converges uniformly, whereas the differentiated function is continuous (see \cite{zorich}, Chapter 17). By formally differentiating the function under the integral sign and omitting the arguments $t$, we obtain
\begin{align}\label{eq:integral}
\frac{1}{\sigma^2}\int_{\R^{Nd}}J_i(\bx)(x^i_k - \mu^i_k)p(\bmu,\bx,\sigma)d\bx.
\end{align}
Given Assumption\r\ref{assum:convex}, $J_i(\bx)(x^i_k - \mu^i_k)p(\bmu,\bx,\sigma)$ is continuous. Thus, it remains to check that the integral of this function converges uniformly with respect to any $\bmu \in \R^{Nd}$. To this end, we can write the Taylor expansion of the function $J_i$ around the point $\bmu(i,k)\in\R^{Nd}$ with the coordinates $\mu(i,k)^i_k = \mu^i_k$ and $\mu(i,k)^j_m=x^j_m$ for any $j\ne i$, $m\ne k$,  in the integral \eqref{eq:integral}:
 \begin{align*}
  &\int_{\R^{Nd}}J_i(\bx)(x^i_k - \mu^i_k)p(\bmu,\bx,\sigma)d\bx \cr
  &= \int_{\R^{Nd}}[J_i(\bmu(i,k)) \cr
  &\quad+ \frac{\partial J_i(\boldsymbol{\eta}(\bx,\bmu))}{\partial x^i_k}(x^i_k - \mu^i_k)](x^i_k - \mu^i_k)p(\bmu,\bx,\sigma)d\bx\cr
  &=\int_{\R^{Nd}}\frac{\partial J_i(\boldsymbol{\eta}(\bx,\bmu))}{\partial x^i_k}(x^i_k - \mu^i_k)^2p(\bmu,\bx,\sigma)d\bx\cr
  &=\int_{\R^{Nd}}\frac{\partial J_i(\boldsymbol{\eta}_1(\by,\bmu))}{\partial x^i_k}(y^i_k)^2 p(\boldsymbol 0,\by,\sigma)d\by,
 \end{align*}
where $\boldsymbol{\eta}(\bx,\bmu)=\bmu(i,k)+\theta(\bx-\bmu(i,k))$, $\theta\in(0,1)$, $\by = \bx-\bmu(i,k)$, ${\boldsymbol{\eta}_1}(\by,\bmu) = \bmu(i,k) +\theta\by$.
The uniform convergence of the integral above follows from the fact\footnote{see the basic sufficient condition using majorant \cite{zorich}, Chapter 17.2.3.} that, under Assumption\r\ref{assum:inftybeh},  $\frac{\partial J_i(\boldsymbol{\eta}_1(\by,\bmu))}{\partial x^i_k}\le l^i_k$ for some positive constant $l^i_k$ and for all $i\in[N]$ and $k\in[d]$. Hence,
\[|\frac{\partial J_i(\boldsymbol{\eta}_1(\by,\bmu))}{\partial x^i_k}(y^i_k)^2 p(\boldsymbol 0,\by,\sigma)|\le h(\by)=l(y^i_k)^2 p(\boldsymbol 0,\by,\sigma),\]
where $\int_{\R^{Nd}}h(\by)d\by<\infty$.
\end{IEEEproof}
Lemma \eqref{lem:sample_grad} shows that the second term inside the projection in \eqref{eq:regpl} is a sample of the gradient of the cost function in mixed strategies. Hence,  algorithm \eqref{eq:regpl}  can be interpreted as a regularized stochastic projection algorithms. To  bound the bias and variance terms of the stochastic projection and consequently establish convergence of the iterates $\bmu(t)$, the parameters $\gamma(t)$, $\sigma(t)$, $\e(t)$ need to satisfy certain assumptions.
\begin{assumption}\label{assum:timestep}
Let $\beta(t)=\gamma(t)\sigma^2(t)$ and choose $\gamma(t)=\frac{1}{t^a}$, $\sigma(t) = \frac{1}{t^b}$ and  $\e(t)=\frac{1}{t^c}$, $a,b,c>0$ respectively, such that 

 a) $\sum_{t=0}^{\infty}\beta(t)=\infty, \quad \lim_{t\to\infty}\e(t)=0,$

 b) $\sum_{t=0}^{\infty}\left(1 + \frac{1}{\beta(t)\e(t)}\right)\frac{|\e(t-1)-\e(t)|^2}{\e^2(t)} < \infty,$
  
 c) $\sum_{t=0}^{\infty}\gamma^2(t)<\infty,$ $\sum_{t=0}^{\infty}\beta(t)\sigma(t) < \infty$,
 
 d) $\lim_{t\to\infty}\sigma(t)=0$, $\sum_{t=0}^{\infty}\beta(t)\e(t)=\infty.$

 \end{assumption}
\begin{theorem}\label{th:main}
Let the players in game $\Gamma(N, \{A_i\}, \{J_i\})$ choose the states $\{\xb^i(t)\}$ at time $t$ according to the normal distribution $\EuScript N(\bmu^i(t),\sigma(t))$, where the mean $\bmu^i(0)$ is arbitrary and $\bmu^i(t)$ is updated as in \eqref{eq:regpl}.  Under Assumptions\r\ref{assum:convex}-\ref{assum:timestep}, as $t\to\infty$, the mean vector $\bmu(t)$ converges almost surely to a Nash equilibrium $\bmu^*=\ab^*$ of the game $\Gamma$ and the joint state $\xb(t)$ converges in probability to $\ab^*$.
\end{theorem}
 \begin{rem}\label{rem:example}
As an example for existence of parameters to satisfy Assumption~\ref{assum:timestep}, let $a=\frac{5}{9}$, $b=\frac{5}{27}$, $c=\frac{1}{27}$.
 \end{rem}

\section{Analysis of the Algorithm}\label{sec:proof}
To prove Theorem\r\ref{th:main} we first prove boundedness of the iterates $\bmu(t)$.  Due to the regularization term $\epsilon(t)$, this is done by analyzing distance of $\bmu(t)$ from the so-called Tikhonov trajectory. Having established this boundedness, we can readily show that the limit of the iterates $\bmu(t)$ exists and satisfies the conditions of a Nash equilibrium of the game $\Gamma(N, \{A_i\}, \{J_i\})$. For  the boundedness and the convergence proofs, we use established results on boundedness (\cite{NH}, Theorem\r2.5.2) and convergence of a sequence of stochastic processes (Lemma 10 (page 49) in \cite{polyak}), respectively. For ease of reference, we provide the statement of (\cite{NH}, Theorem\r2.5.2) and  (Lemma 10 (page 49) in \cite{polyak} ) in the appendix. 

\subsection{Boundedness of the Algorithm Iterates}
We first show that  algorithm \eqref{eq:regpl} falls under the framework of well-studied Robbins-Monro stochastic approximations procedures  \cite{Borkar} with an additional regularization $\epsilon(t)$. Next, leveraging this analogy and results on stability of discrete-time Markov processes (\cite{NH}, Theorem\r2.5.2) applied to the sequence $\bmu(t)$ we prove boundedness of the iterates.

Using the notation $\Mb_i (\cdot)=(M_{i,1}(\cdot), \ldots, M_{i,d}(\cdot))$,
we can rewrite the algorithm step in \eqref{eq:regpl} in the following form:
 \begin{align}
 \label{eq:pbavmu}
&\bmu^i(t+1) =\Proj_{\Ab_i}[\bmu^i(t) -\gamma(t)\sigma^2(t)\cr
&\times\big(\Mb_i(\bmu(t)) +\Qb_i(\bmu(t),\sigma(t))+\Rb_i(\bmu(t),\xb(t),\sigma(t))\cr
&\qquad\qquad\qquad\qquad\qquad\qquad\qquad+\epsilon(t))\bmu^i(t)\big),
\end{align}
for all $i\in[N]$
 \begin{align*}
&\Qb_i(\bmu(t),\sigma(t)) =\tilde{\Mb}_i (\bmu(t),\sigma(t)) -\Mb_i(\bmu(t)),\cr
&\Rb_i(\xb(t),\bmu(t),\sigma(t)) = \boldsymbol F_i(\xb(t),\bmu(t), \sigma(t)) - \tilde{\Mb}_i (\bmu(t),\sigma(t)), \cr
&\boldsymbol F_i(\xb(t),\bmu(t), \sigma(t)) ={\hat J}_i(t)\frac{\xb^i(t) -\bmu^i(t)}{\sigma^2(t)} ,
\end{align*}
and $\tilde{\Mb}_i (\cdot)=(\tilde M_{i,1}(\cdot), \ldots, \tilde M_{i,d}(\cdot))^{\top}$
is the $d$-dimensional mapping with the following elements:
\begin{align}\label{eq:mapp2}
\tilde M_{i,k} (\bmu(t),\sigma(t))=\frac{\partial {\tilde J_i(\bmu(t), \sigma(t))}}{\partial \mu^i_k}, \mbox{ for $k\in[d]$}.
\end{align}
The vector $\Mb(\bmu(t))=(\Mb_1(\bmu(t)), \ldots, \Mb_N(\bmu(t)))$
corresponds to the gradient term in stochastic approximation procedures, whereas
 \begin{align*}
  \Qb(\bmu(t),\sigma(t)) = (\Qb_1(\bmu(t),\sigma(t)),\ldots,&\Qb_N(\bmu(t),\sigma(t)))
 \end{align*}
is a disturbance of the gradient term. Finally,
 \begin{align*}
  \Rb(\xb(t), \bmu(t),\sigma(t)) = (&\Rb_1(\xb(t),\bmu(t),\sigma(t)),\ldots,\cr
&\Rb_N(\xb(t), \bmu(t),\sigma(t)))
 \end{align*}
is a martingale difference, namely, according to \eqref{eq:mathexp},
\begin{align}
\label{eq:mathexp2}
 &\Rb_i(\xb(t),\bmu(t),\sigma(t)) = \boldsymbol F_i(\xb(t),\bmu(t),\sigma(t)) \\
\nonumber
 &- \E_{\xb(t)}\{\boldsymbol F_i(\xb(t),\bmu(t),\sigma(t))\},\; \;i\in[N].
\end{align}
To ensure boundedness of $\bmu(t)$ (Lemma \ref{lem:boundedvec})  we  bound the martingale term above (see Inequality \eqref{eq:Rineq1}). To bound  the disturbance of the gradients $  \Qb(\bmu(t),\sigma(t))$ (see Equation \eqref{eq:Qterm1}), we observe that the mapping $\tilde{\Mb}_i (\bmu(t))$ evaluated at $\bmu(t)$ is equivalent to the  game mapping in mixed strategies 
(please see Appendix for the proof of this observation).
That is,
\begin{align}\label{eq:gradmix}
 \tilde{\Mb}_i (&\bmu(t))=\int_{\mathbb R^{Nd}}{\Mb_i} (\bx)p(\bmu(t),\bx)d\bx.
\end{align}
In contrast to stochastic approximation algorithms and the proof in \cite{TAC2018tat_kam}, we have an addition term $\epsilon(t) \bmu(t)$ to be able to address merely monotone game mappings. As such, to bound $\bmu(t)$ we also relate the variations of the sequence  $\bmu(t)$ to those of the \emph{Tikhonov sequence} defined below. 
Let $\by(t) = (\by^1(t), \ldots, \by^N(t))$ denote the solution of the variational inequality $VI(\Ab,\Mb(\by) + \e(t)\by)$, namely 
\begin{align}\label{eq:Tikhonov}
 \by(t)\in SOL(\Ab,\Mb(\by) + \e(t)\by).
\end{align}
The sequence $\{\by(t)\}$  is known  as the Tikhonov sequence and enjoys the following two important properties. 
\begin{theorem}\label{th:Tikhonov}{(Theorem 12.2.3 in \cite{FaccPang1})}
Under Assumptions~\ref{assum:CG_grad}, \ref{assum:existence}, and \ref{assum:Lipschitz},  $\by(t)$ defined in \eqref{eq:Tikhonov} exists and is unique for each $t$. Moreover, for $\epsilon(t) \downarrow 0$,  $\by(t)$ is uniformly bounded and converges to the least norm solution of $VI(\Ab,\Mb)$.
\end{theorem}
\begin{lem}\label{lem:Koshal}(Lemma~3 in \cite{koshal2010single})
Under Assumption~\ref{assum:CG_grad}
\[\|\by(t)-\by(t-1)\|\le M_{\by}\frac{|\e(t-1)-\e(t)|}{\e(t)}, \quad \forall t \geq 1,\]
where $M_{\by}$ is a uniform bound on the norm of the Tikhonov sequence, i.e. $\|\by(t)\|\le M_{\by}$ for all $t\ge 0$.
\end{lem}

With the results above in place, we  connect the squared distance $\|\bmu - \by(t)\|^2$ to the squared distance $\|\bmu - \by(t-1)\|^2$ for any $\bmu\in \Ab$ and $t\ge 1$.
Due to the triangle inequality, 
\begin{align}\label{eq:triangle}
 \|\bmu - \by(t)\|&\le \|\bmu - \by(t-1)\| + \|\by(t-1)- \by(t)\| \\
 \nonumber
 &\le \|\bmu - \by(t-1)\| +M_{\by}\frac{|\e(t-1)-\e(t)|}{\e(t)},
\end{align}
where in the last inequality we used Lemma\r\ref{lem:Koshal}. Hence, by taking into account that for any $a, b\in \R$ and  $\theta>0$
\[2ab\le \theta a^2 + \frac{b^2}{\theta},\]
we conclude from \eqref{eq:triangle} that for any $\theta>0$
\begin{align}\label{eq:connection}
 \|\bmu - \by(t)\|^2\le &(1+\theta)\|\bmu - \by(t-1)\|^2  \cr
 & +\left(1 + \frac{1}{\theta}\right)M^2_{\by}\frac{|\e(t-1)-\e(t)|^2}{\e^2(t)}.
\end{align}
The above bound serves as the main new inequality in order to show almost-sure boundedness of $\| \bmu(t) \|$ in comparison to non-regularized stochastic gradient procedures. 

\begin{lem}\label{lem:boundedvec}
 Let Assumptions\r\ref{assum:CG_grad}-\ref{assum:timestep} hold in $\Gamma(N, \{A_i\}, \{J_i\})$  and $\bmu(t)$ be the vector updated in the run of the payoff-based algorithm \eqref{eq:pbavmu}. Then, $\Pr\{\sup_{t\ge 0}\|\bmu(t)\|< \infty\}=1$.
\end{lem}

In the following, for simplicity in notation, we omit the argument $\sigma(t)$ in the terms  $\tilde{\Mb}$, $\Qb$, and $\Rb$. In certain derivations, for the same reason we omit the time parameter $t$ as well. 

\begin{IEEEproof}
Define $V(t,\bmu) = \|\bmu-\by(t-1)\|^2$, where $\by(t)$ is the Tikhonov sequence defined by \eqref{eq:Tikhonov}. We consider the generating operator of the Markov process $\bmu(t)$
\begin{align*}
LV(t,\bmu)=E[V(t+1, \bmu(t+1))\mid \bmu(t)=\bmu]-V(t,\bmu),
\end{align*}
and aim to show that $LV(t,\bmu)$ satisfies the following decay 
\begin{align}
\label{eq:decay}
LV(t,\bmu)\le -\alpha(t+1)\psi(\bmu) + \phi(t)(1+V(t,\bmu)),
\end{align}
where $\psi\ge 0$ on $ \R^{Nd}$, $\phi(t)>0$, $\forall t$, $\sum_{t=0}^{\infty}\phi(t)<\infty$, $\alpha(t)>0$, $\sum_{t=0}^{\infty} \alpha(t)= \infty$. This enables us to apply Theorem\r2.5.2 in \cite{NH} to directly conclude almost sure boundedness of $\bmu(t)$.

Let us bound the growth of $V(t+1,\bmu)$ in terms of $V(t, \bmu)$. Let $\theta = \beta(t)\e(t)$ in \eqref{eq:connection}. From Assumption~\ref{assum:timestep} b), $\left(1 + \frac{1}{\beta(t)\e(t)}\right)\frac{|\e(t-1)-\e(t)|^2}{\e^2(t)} \rightarrow 0$ as $t\to\infty$. Hence,  $\forall \bmu\in\Ab$
   \begin{align}\label{eq:connection1}
    V(t+1,\bmu) =& \|\bmu-\by(t)\|^2\\ \nonumber
    \le& (1+\beta(t)\e(t))\|\bmu - \by(t-1)\|^2  \cr
 & +\left(1 + \frac{1}{\beta(t)\e(t)}\right)M^2_{\by}\frac{|\e(t-1)-\e(t)|^2}{\e^2(t)}\\\nonumber
 =& O(1 + \|\bmu-\by(t-1)\|^2)= O(1 + V(t,\bmu)).
   \end{align}

From the procedure for the update of $\bmu(t)$, the non-expansion property of the projection operator, the fact that $\by(t)$ belongs to $SOL(\Ab,\Mb(\by) + \e(t)\by)$, namely, that $ \forall i\in[N]$
\[\by^i(t) = \Proj_{A_i}[\by^i(t) - \beta(t)(\Mb_i(\by(t)) + \epsilon(t)\by^i(t)],\]
we obtain that for any $i\in[N]$
\begin{align}\label{eq:nonexp}
 \|&\bmu^i(t+1)-\by^i(t)\|^2 \cr
 &\le \|\bmu^i(t)-\by^i(t) -\beta(t)\big[\epsilon(t)(\bmu^i(t)-\by^i(t))\cr
 &+(\Mb_i(\bmu(t)) - \Mb_i(\by(t)) +\Qb_i(\bmu(t))+\Rb_i(\xb(t),\bmu(t))\big]\|^2\cr
 & = \|\bmu^i(t)-\by^i(t)\|^2 \cr
 &\qquad- 2\beta(t)(\Mb_i(\bmu(t))- \Mb_i(\by(t)), \bmu^i(t)-\by^i(t)) \cr
 &\qquad - 2\beta(t)\epsilon(t)(\bmu^i(t)-\by^i(t), \bmu^i(t)-\by^i(t))\cr
 &\qquad-2\beta(t)(\Qb_i(\bmu(t))+\Rb_i(\xb(t),\bmu(t)), \bmu^i(t)-\by^i(t)) \cr
 &\qquad + \beta^2(t)\|\Gb_i(\xb(t),\bmu(t))\|^2,
\end{align}
where, for ease of notation, we have defined
\begin{align}\label{eq:G_1}
 \Gb_i(\xb(t),\bmu(t)) = &\epsilon(t)(\bmu^i(t)-\by^i(t)) \cr
 &+ \Mb_i(\bmu(t)) - \Mb_i(\by(t))\cr
 &+\Qb_i(\bmu(t))+\Rb_i(\xb(t),\bmu(t)).
\end{align}
Our goal is to bound $\E\{\|\bmu^i(t+1)-\by^i(t)\|^2|\bmu(t) = \bmu\}$ above, and use this bound in constructing Inequality \eqref{eq:decay}. As such, we expand $\Gb_i$ as below and bound the terms in the expansion.
\begin{align}\label{eq:G_1_norm}
 &\|\Gb_i(\xb(t),\bmu(t))\|^2 = \epsilon^2(t)\|\bmu^i(t)-\by^i(t)\|^2 \cr
 &+ \|\Mb_i(\bmu(t)) - \Mb_i(\by(t))\|^2\cr
 &+ \|\Qb_i(\bmu(t))\|^2+\|\Rb_i(\xb(t),\bmu(t))\|^2\cr
 & + 2(\Qb_i(\bmu(t)),\Rb_i(\xb(t),\bmu(t)))\cr
 & + 2\e(t)(\Mb_i(\bmu(t)) - \Mb_i(\by(t)),\bmu^i(t)-\by^i(t))\cr
 & +2(\e(t)(\bmu^i(t)-\by^i(t)) + \Mb_i(\bmu(t)) - \Mb_i(\by(t)),\cr
 &\qquad\qquad\qquad\qquad\Qb_i(\bmu(t))+\Rb_i(\xb(t),\bmu(t))),
\end{align}
Due to Assumption~\ref{assum:Lipschitz}, we conclude that 
\begin{align}\label{eq:Mterm1}\nonumber
 &\|\Mb_i(\bmu) - \Mb_i(\by(t))\|^2 \le L^2_i\|\bmu-\by(t)\|^2 = O(V(t+1,\bmu)) \\
  &\le O(1 + V(t,\bmu)),\\
\label{eq:Mterm2}
&(\Mb_i(\bmu)-\Mb_i(\by(t)),\bmu^i-\by^i(t))\cr
&\le\|\Mb_i(\bmu)-\Mb_i(\by(t))\|\|\bmu^i-\by^i(t)\|\cr
&\le O(1+V(t+1,\bmu))\le O(1 + V(t,\bmu)),
\end{align}
where in the last inequalities in \eqref{eq:Mterm1}-\eqref{eq:Mterm2} we used \eqref{eq:connection1}. Let us analyze the terms containing the disturbance of gradient, namely $\Qb_i$, in Equation \eqref{eq:G_1_norm}. Since $\Qb_i(\bmu(t))=\tilde{\Mb}_i(\bmu(t)) - \Mb_i(\bmu(t))$, due to Assumption~\ref{assum:CG_grad} and Equation \eqref{eq:gradmix}, we obtain
\begin{align}\label{eq:Qterm1}
 \|\Qb_i(\bmu)\|&=\|\int_{\R^{Nd}}[\Mb_i(\bx)-\Mb_i(\bmu)]p(\bmu,\bx)d\bx\| \cr
 &\le \int_{\R^{Nd}}\|\Mb_i(\bx) - \Mb_i(\bmu)\| p(\bmu,\bx) d\bx\cr
 &\le \int_{\R^{Nd}} L_i \|\bx - \bmu\| p(\bmu,\bx) d\bx \cr
 \nonumber &\le \int_{\R^{Nd}} L_i \left(\sum_{i=1}^{N}\sum_{k=1}^{d}|x^i_k - \mu^i_k|\right) p(\bmu,\bx) d\bx\\
 &= O(\sum_{i=1}^{N}\sigma),
 \end{align}
where the last equality is due to the fact that the first central absolute moment of a random variable with a normal distribution $\EuScript N(\mu,\sigma)$ is $O(\sigma)$.
The estimation above and \eqref{eq:connection1} imply, in particular, that for any $\bmu\in \Ab$
\begin{align}\label{eq:Qterm2}
&\|\Qb_i(\bmu)\|\|\bmu^i-\by^i(t)\|\le O(\sum_{i=1}^{N}\sigma)(1+V(t,\bmu))\\
\nonumber
&\|\Qb_i(\bmu)\|\|M_i(\bmu)-M_i(\by(t))\|\le L_i\|\Qb_i(\bmu)\|\|\bmu-\by(t)\|\\
\label{eq:Qterm3}
&\le O(\sum_{i=1}^{N}\sigma)(1+V(t,\bmu)).
\end{align}
Finally, we bound the martingale term $\|\Rb_i(\xb(t),\bmu(t))\|^2$.
 \begin{align}  
 \label{eq:Rineq1}
 \nonumber
  &\E\{\|\Rb_i(\xb(t),\bmu(t))\|^2|\bmu(t)=\bmu\}\\ \nonumber
  &\le\sum_{k=1}^{d}\int_{\mathbb R^{Nd}}{J_i}^2(\bx)\frac{(x^i_k - \mu^i_k(t))^2}{\sigma^4(t)} p(\bmu,\bx)d\bx\\
&\le \frac{f_i(\bmu, \sigma(t))}{\sigma^4(t)} \le \frac{O(1+V(t,\bmu))}{\sigma^4(t)} ,
  \end{align}
where the first inequality is due to the fact that $\E(\xi-\E\xi)^2\le\E\xi^2$ and taking into account \eqref{eq:mathexp2}, the second inequality is due to Assumption~\ref{assum:inftybeh}, with $f_i(\bmu, \sigma(t))$ being a quadratic function of $\bmu$ and $\sigma(t)$, $i\in[N]$.
Bringing the inequalities \eqref{eq:Mterm1}-\eqref{eq:Rineq1} in the inequality \eqref{eq:nonexp}, taking into account \eqref{eq:connection1}, the Cauchi-Schwarz inequality, 
and the martingale properties in \eqref{eq:mathexp2} of $\Rb_i$, $i\in[N]$, we get 
 \begin{align}\label{eq:mu_part}
  \E\{\|&\bmu^i(t+1)-\by^i(t)\|^2|\bmu(t) = \bmu\} \cr
 &\le (1-2\beta(t)\epsilon(t))\|\bmu^i-\by^i(t)\|^2 \cr
 &\quad- 2\beta(t)(\Mb_i(\bmu)- \Mb_i(\by(t)), \bmu^i-\by^i(t)) \cr
  &\quad-2\beta(t)(\Qb_i(\bmu), \bmu^i-\by^i(t)) \cr
 &\quad + \beta^2(t)\E\{\|\Gb_i(\xb(t),\bmu)\|^2|\bmu(t)=\bmu\}\cr
 &\le (1-2\beta(t)\epsilon(t))\|\bmu^i-\by^i(t)\|^2 \cr
 &\quad- 2\beta(t)(\Mb_i(\bmu)- \Mb_i(\by(t)), \bmu^i-\by^i(t)) \cr
 &\quad+2\beta(t)O(\sum_{i=1}^{N}\sigma(t))(1+V(t,\bmu))\cr
 &\quad+O(\gamma^2(t))(1+V(t,\bmu)),
 \end{align}
where in the last inequality we used the fact that $\e(t)\to 0$ (Assumption~\ref{assum:timestep} a)), $\gamma(t)\to 0$, and $\sigma(t)\to 0$ 
for all $i\in[N]$ as $t\to\infty$ (Assumption~\ref{assum:timestep} c), d)).  Thus, taking into account Assumption\r\ref{assum:timestep} c), d) and \eqref{eq:mu_part},  we obtain
\begin{align}\label{eq:LV}
\E[&\|\bmu(t+1)-\by(t)\|^2|\bmu(t)=\bmu]\cr
=&\sum_{i=1}^N\E[\|\bmu^i(t+1)-\by^i(t)\|^2|\bmu(t)=\bmu]\cr
\le&(1-2\e(t)\beta(t))\|\bmu - \by(t)\|^2\cr
&-2\beta(t)(\Mb(\bmu)-\Mb(\by(t)), \bmu-\by(t))\cr
& + O(\beta(t)\sigma(t)+\gamma^2(t))(1+V(t,\bmu)).
\end{align}
Using the first inequality in \eqref{eq:connection1}, we get 
\begin{align}\label{eq:LV0}
 &LV(t,\bmu) \cr
 &= \E[\|\bmu(t+1)-\by(t)\|^2|\bmu(t)=\bmu] - \|\bmu-\by(t-1)\|^2\cr
 &\le \E[\|\bmu(t+1)-\by(t)\|^2|\bmu(t)=\bmu]   \cr
 &\quad-\frac{\|\bmu-\by(t)\|^2 - M_{\by}^2\left(1 + \frac{1}{\beta(t)\e(t)}\right)\frac{|\e(t-1)-\e(t)|^2}{\e^2(t)}}{1+\beta(t)\e(t)}.
\end{align}
We conclude from \eqref{eq:LV0} and \eqref{eq:LV} that 
\begin{align}\label{eq:LV1}
\nonumber
 &LV(t,\bmu) \le \left(1-2\e(t)\beta(t)-\frac{1}{1+\e(t)\beta(t)}\right)\|\bmu - \by(t)\|^2\\ \nonumber
 &-2\beta(t)(\Mb(\bmu)-\Mb(\by(t)), \bmu-\by(t))\\ \nonumber 
 &\qquad\qquad\qquad + h(t)(1 + V(t,\bmu))\\ \nonumber
 &\le -2\beta(t)(\Mb(\bmu)-\Mb(\by(t)), \bmu-\by(t)) \\
 &\qquad\qquad\qquad + h(t)(1 + V(t,\bmu)),
\end{align}
where 
\begin{align}\label{eq:ht}
 h(t) = & O(\beta(t)\sigma(t)+\gamma^2(t))\cr
 &+ O\left(\left(1 + \frac{1}{\beta(t)\e(t)}\right)\frac{|\e(t-1)-\e(t)|^2}{\e^2(t)}\right),
\end{align}
and the second inequality above is due to the fact that 
\[(1-2\e(t)\beta(t))(1+\e(t)\beta(t))\le 1.\]
According to Assumption\r\ref{assum:timestep} b)-c), $\sum_{t=0}^{\infty}h(t)<\infty$. Furthermore, from Assumption~\ref{assum:timestep} a) $\sum_{t=0}^{\infty}\beta(t)=\infty$. Taking into account this, \eqref{eq:LV1},  and monotonicity of $\Mb$ implying 
\begin{align}\label{eq:monotone}
 (\Mb(\bmu)-\Mb(\by(t)),\bmu-\by(t))\ge0, \; \forall t,\, \forall \bmu\in\Ab,
\end{align}
we conclude that   $LV(t,\bmu)$ satisfies the decay needed for the application of Theorem\r2.5.2 in \cite{NH} and consequently, $\bmu(t)$ is finite almost surely for any $t\in\Z_{+}$ irrespective of $\bmu(0)$.
\end{IEEEproof}

\subsection{Convergence of the Algorithm}
Fortunately, the derivations in the previous section in proving boundedness of the iterates can be used to also prove convergence of the algorithm. In particular, we use Inequality \eqref{eq:LV}, which bounds  the decay of the sequence $\E[\|\bmu(t+1)-\by(t)\|^2|\bmu(t)]$ in terms of $\|\bmu - \by(t)\|^2$. We can show that this decay satisfies the conditions for applying Lemma 10 in \cite{polyak}. From this, it can  readily be inferred that random variables $\|\bmu(t)-\by(t-1)\|$ converge to zero. In essence, the approach is similar to showing that $V(t,\mu)$ serves as a stochastic Lyapunov function for the sequence of random variables.
\begin{IEEEproof} (of Theorem~\ref{th:main})
First, rewrite \eqref{eq:LV} as follows:
\begin{align}\label{eq:LV2}
&\E[\|\bmu(t+1)-\by(t)\|^2|\EuScript F_t]\cr
\le&(1-2\e(t)\beta(t))\|\bmu(t) - \by(t)\|^2\cr
& + O(\gamma^2(t)+\beta(t)\sigma(t))(1+V(t,\bmu(t)))\cr
\le&(1-2\e(t)\beta(t))\|\bmu(t) - \by(t)\|^2 + O(\gamma^2(t)+\beta(t)\sigma(t))\cr
\le&(1-2\e(t)\beta(t))(1+\e(t)\beta(t))\|\bmu(t) - \by(t-1)\|^2+ O(h(t))\cr
\le&(1-\e(t)\beta(t))\|\bmu(t) - \by(t-1)\|^2+ O(h(t)),
\end{align}
where $\EuScript F_t$ is the $\sigma$-algebra generated by the random variables $\{\xb(k),\bmu(k)\}_{k=0}^t$ and $h(t)$ is defined in \eqref{eq:ht}. In \eqref{eq:LV2} 
to get the first inequality we used \eqref{eq:monotone}, to get the second inequality we used Lemma~\ref{lem:boundedvec}, namely the fact that $\bmu(t)$ is almost surely bounded for all $t\in\Z_{+}$, to get the third inequality we used 
\eqref{eq:connection1}, and to get the last inequality we used the fact that $(1-2\e(t)\beta(t))(1+\e(t)\beta(t))<(1-\e(t)\beta(t))$.

From Assumption~\ref{assum:timestep}, and the choices of $\gamma(t)$, $\sigma(t)$,  $\e(t)$, we get $O(h(t))=\frac{1}{t^l}$, $\e(t)\beta(t)=\frac{1}{t^m}$, with $l>1$, $m\le 1$. Thus,
\[\lim_{t\to\infty}\frac{O(h(t))}{\e(t)\beta(t)}= 0.\]
Assumption~\ref{assum:timestep} d), the fact that $\sum_{t=0}^\infty h(t) < \infty$ and the above result in the decay \eqref{eq:LV2} imply that we can apply Lemma 10 in \cite{polyak} to the sequence $\|\bmu(t+1)-\by(t)\|^2$ to conclude its almost sure convergence to $0$ as $t\to\infty$. Next, by taking into account Theorem~\ref{th:Tikhonov} and Theorem~\ref{th:VINE}, we obtain that
$$\Pr\{\lim_{t\to\infty}\bmu(t)=\ab^*\}=1,$$
where $\ab^*$ is the least norm Nash equilibrium in the game $\Gamma(N, \{A_i\}, \{J_i\})$. Finally, Assumption\r\ref{assum:timestep} implies that $\lim_{t\to\infty}\sigma(t)=0$.
Taking into account that $\xb(t)\sim\EuScript N(\bmu(t),\sigma(t))$, we conclude that $\xb(t)$ converges weakly to a Nash equilibrium $\ab^*=\bmu^*$. Moreover, according to  Portmanteau Lemma \cite{portlem}, this convergence is also in probability.

\end{IEEEproof}

\section{Simulation Results}
As noted in the introduction, the work \cite{grammatico2018comments} provides a counterexample showing that the class of gradient-based procedures proposed in \cite{ZhuFrazzoli} and \cite{TAC2018tat_kam} fail to converge to a Nash equilibrium, if the game mapping is merely monotone. In this section, we demonstrate that the inclusion of the Tikhonov regularization term in algorithm \cite{TAC2018tat_kam} rectifies this issue. In particular, the payoff-based algorithm proposed here converges to the Nash equilibrium in the game under consideration. 

\begin{figure}[!htb]
	\centering
	\includegraphics[width=0.45\textwidth]{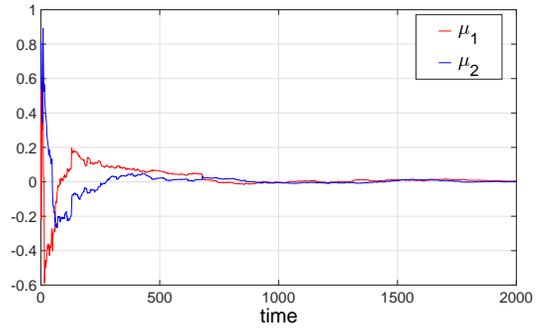}
        \caption{\label{fig:alg_run}The mean values for $\mu_1$ and $\mu_2$ based on Procedure \eqref{eq:regpl}.}
        \end{figure}

Following the discussion in \cite{grammatico2018comments}, we consider the game with $2$ players, whose action sets are $1$-dimensional sets $A_1=A_2=[-1,1]$ and the cost functions are $J_1(a_1,a_2) = a_1a_2$ and $J_2(a_1,a_2) = -a_1a_2$ respectively. It can be verified that the game mapping $M(a_1,a_2) = (a_2,-a_1)$ is monotone and the unique Nash equilibrium in this game is $\ab^*=(0,0)$. By implementing the payoff-based algorithm \eqref{eq:pbavmu} with randomly chosen initial values $\mu^1(0)$ and $\mu^2(0)$ and the parameters $\gamma(t)$, $\sigma(t)$, and $\e(t)$ set up according to Remark~\ref{rem:example}, we obtain the updates for the mean values $\mu^1(t)$ and $\mu^2(t)$ of the players, presented in  Figure~\ref{fig:alg_run}. As we can see, the procedure ensures the means arrive at a sufficiently small neighborhood of the Nash equilibrium after approximately $900$ iterations and continue approaching it in its further run.  

\section{Conclusions}
We proposed a payoff-based algorithm for learning Nash equilibria in convex games with monotone game mappings. Our algorithm relied on a suitable regularization to handle monotonicity. The convergence proof relied on the analysis of the Tikhonov sequence related to the regularization and well-established results on boundedness and convergence of stochastic processes. Our current work addresses establishing convergence rate of the algorithm under suitable assumptions. 
\bibliographystyle{plainurl}
\bibliography{../CentralPath/TAC_ArxivVersion}

\appendix

\subsection{Supporting Theorems}
Let $\{\zbx(t)\}_t$, $t\in \Z_+$, be a discrete-time Markov process on some state space $E\subseteq \R^d$, namely $\zbx(t)=\zbx(t,\omega):\Z_+\times\Omega\to E$, where $\Omega$ is the sample space of the probability space on which the process $\zbx(t)$ is defined. The transition function of this chain, namely $\Pr\{\zbx(t+1)\in\Gamma| \zbx(t)=\zbx\}$, is denoted by $P(t,\zbx,t+1,\Gamma)$, $\Gamma\subseteq E$.

\begin{definition}\label{def:def1}
The operator $L$ defined on the set of measurable functions $V:\Z_+\times E\to \R$, $\zbx\in E$, by
\begin{align*}
LV(t,\zbx)&=\int{P(t,\zbx,t+1,dy)[V(t+1,y)-V(t,\zbx)]}\cr
&=E[V(t+1,\zbx(t+1))\mid \zbx(t)=\zbx]-V(t,\zbx),
\end{align*}
is called a \emph{generating operator} of a Markov process $\{\zbx(t)\}_t$.
\end{definition}
Next, we formulate the following theorem for discrete-time Markov processes, which is proven in \cite{NH}, Theorem\r2.5.2.

\begin{theorem}\label{th:finiteness}
  Consider a Markov process $\{\zbx(t)\}_t$ and suppose that there exists a function $V(t,\zbx)\ge 0$ such that $\inf_{t\ge0}V(t,\zbx)\to\infty$ as $\|\zbx\|\to\infty$ and
  \[LV(t,\zbx)\le -\alpha(t+1)\psi(t,\zbx) + f(t)(1+V(t,\zbx)),\]
   where $\psi\ge 0$ on $\R \times \R^d$, $f(t)>0$, $\sum_{t=0}^{\infty}f(t)<\infty$. Let $\alpha(t)$ be such that $\alpha(t)>0$, $\sum_{t=0}^{\infty} \alpha(t)= \infty$.
   Then, almost surely $\sup_{t\ge 0}\|\zbx(t,\omega)\| = R(\omega)< \infty$.
\end{theorem}

The following result related to the convergence of the stochastic process is proven in Lemma 10 (page 49) in \cite{polyak}. 
\begin{theorem}\label{th:polyak_lem11}
   Let $v_0, \ldots, v_k$ be a sequence of random variables, $v_k\ge 0$, $\E v_0<\infty$ and let
   \[\E \{v_{k+1}| \EuScript F_k\} \le (1-\alpha_k)v_k + \beta_k,\]
   where $\EuScript F_k$ is the $\sigma$-algebra generated by the random variables $\{v_0,\ldots,v_k\}$, $0<\alpha_k<1$, $\sum_{k=0}^{\infty}\alpha_k = \infty$, 
   $\beta_k\ge 0$, $\sum_{k=0}^{\infty}\beta_k < \infty$, $\lim_{k\to\infty}\frac{\beta_k}{\alpha_k}=0$. 
   Then $v_k\to 0$ almost surely, $\E v_k\to 0$ as $k\to \infty$.
\end{theorem}

%

\subsection{Verification of Equation \eqref{eq:gradmix}}
We will show that the mapping $\tilde{\Mb}_i(\bmu(t), \sigma(t))$ (see \eqref{eq:mapp2}) evaluated at $\bmu(t)$ is equivalent to the extended game mapping:
 \begin{align*}
 \tilde{\Mb}_i (&\bmu(t))=\int_{\mathbb R^{Nd}}{\Mb_i} (\bx)p(\bmu(t),\bx)d\bx.
\end{align*}
Note that for simplicity in notation, we drop the dependence on $\sigma(t)$ and on $t$. Now, using the notations
\begin{align*}
&\mu^i_{-k}=(\mu^i_1,\ldots,\mu^{i}_{k-1},\mu^{i}_{k-1},\ldots\mu^{i}_{d})\in\R^{d-1},\\
&x^i_{-k}=(x^i_1,\ldots,x^{i}_{k-1},x^{i}_{k-1},\ldots x^{i}_{d})\in\R^{d-1},\\
&p(\mu^i_{-k},x^i_{-k})=\frac{1}{(\sqrt{2\pi}\sigma_i)^{d-1}}\exp\left\{-\sum_{j\ne k}\frac{(x^i_j-\mu^i_j)^2}{2\sigma_i^2}\right\}\\
&p(\bmu^{-i},\bx^{-i})=\prod_{j\ne i, j=1}^N\frac{1}{(\sqrt{2\pi}\sigma_j)^{d}}\exp\left\{-\sum_{k=1}^d\frac{(x^j_k-\mu^j_k)^2}{2\sigma_j^2}\right\},
\end{align*}
we have that for any $i\in[N]$, $k\in[d]$, $\tilde{M}_{i,k}(\be)$
\begin{align}\label{eq:diffpotfun}
\allowdisplaybreaks
&\tilde{M}_{i,k}(\bmu) = \frac{\partial {\tilde J_i(\bmu(t), \sigma(t))}}{\partial \mu^i_k}\cr
&= \frac{1}{\sigma_i^2}\int_{\R^{Nd}}J_i(\bx)(x^i_k - \mu^i_k)p(\bmu,\bx)d\bx\cr
&= -\int_{\R^{Nd}}J_i(\bx) p(\mu^i_{-k},x^i_{-k}) p(\bmu^{-i},\bx^{-i})\frac{1}{\sqrt{2\pi}\sigma_i} \cr
&\qquad\qquad\qquad\qquad\qquad\qquad\qquad \times d\left(e^{-\frac{(x^i_k-\mu^i_k)^2}{2\sigma_i^2}}\right) d\bx^{-i}\cr
 &=-\int_{\R^{Nd-1}}\left(J_i(\bx)e^{-\frac{(x_k^i-\mu_k^i)^2}{2\sigma_i^2}}\right)\bigg|_{-\infty(x_k^i)}^{\infty(x_k^i)}\cr
&\qquad\qquad\qquad \times p(\mu^i_{-k},x^i_{-k}) p(\bmu^{-i},\bx^{-i})\frac{1}{\sqrt{2\pi}\sigma_i}d\bx^{-i}\cr
 &\qquad\qquad\qquad\qquad+\int_{\R^{Nd}}\frac{\partial J_i(\bx)}{\partial x^i_k}p(\bmu,\bx)d\bx\cr
 &=\int_{\R^{Nd}}\frac{\partial J_i(\bx)}{\partial x^i_k}p(\bmu,\bx)d\bx.
 \end{align}
In the above, for the second equality, we used Lemma \eqref{lem:sample_grad} to enable differentiation under the integral and for the last equality, we used the fact that according to Assumption~\ref{assum:inftybeh},
\begin{align*}
\lim_{x_k^i\to\infty(-\infty)}J_i(\bx)e^{-\frac{(x_k^i-\mu_k^i)^2}{2\sigma_i^2}}=0,
\end{align*}
 for any fixed $\mu_k^i$, $\bx^{-i}$. Now, by definition of $\Mb_i(\bx)$, we have that 
 \begin{align*}
 \int_{\R^{Nd}}\frac{\partial J_i(\bx)}{\partial x^i_k}p(\bmu,\bx)d\bx = \int_{\mathbb R^{Nd}}{\Mb_i} (\bx)p(\bmu(t),\bx)d\bx,
 \end{align*}
as desired. 

\end{document}